\documentclass[a4paper, 11pt]{article}

\usepackage{graphicx}
\usepackage{xcolor}
\usepackage{slashed}
\usepackage{ amssymb }
\usepackage{mathtools}
\usepackage{latexsym}
\usepackage{textgreek}
\usepackage{verbatim}
 \usepackage[utf8]{inputenc}
 \usepackage{amsmath}
\usepackage{amsfonts}
\usepackage{verbatim}
\usepackage{graphicx}
\usepackage{subfigure}
\usepackage{amssymb}
\usepackage[T1]{fontenc}
\usepackage{slashed}
\usepackage{color}
\usepackage[numbers,sort&compress]{natbib}
\usepackage[linktocpage=true]{hyperref}

\def\beq{\begin{eqnarray}}
\def\eeq{\end{eqnarray}}

\def\({\left(}
\def\){\right)}

\newcommand{\be}{\begin{equation}}
\newcommand{\ee}{\end{equation}}

\def\ea{\end{eqnarray}}
\def\ba{\begin{eqnarray}}

\def\beq{\begin{eqnarray}}
\def\eeq{\end{eqnarray}}

\def\({\left(}
\def\){\right)}

\def\lsim{\mathrel{\rlap{\lower3pt\hbox{\hskip0pt$\sim$}}
     \raise1pt\hbox{$<$}}}         
\def\gsim{\mathrel{\rlap{\lower4pt\hbox{\hskip1pt$\sim$}}
     \raise1pt\hbox{$>$}}}         

\def\lsim{\mathrel{\rlap{\lower3pt\hbox{\hskip0pt$\sim$}}
     \raise1pt\hbox{$<$}}}         
\def\gsim{\mathrel{\rlap{\lower4pt\hbox{\hskip1pt$\sim$}}
     \raise1pt\hbox{$>$}}}         

\begin{document}

\begin{center}
{\Large \bf{Thermalization, Fragmentation and Tidal Disruption:\\The Complex Galactic Dynamics of Dark Matter Superfluidity}}

 \vspace{1truecm}
\thispagestyle{empty} \centerline{\large  {Lasha  Berezhiani~$^{1}$, Giordano Cintia~$^{1,2}$ and Justin Khoury~$^3$}
}

 \textit{$^1$Max-Planck-Institut f\"ur Physik, F\"ohringer Ring 6, 80805 M\"unchen, Germany\\
 \vskip 5pt
$^2$ Arnold Sommerfeld Center, Ludwig-Maximilians-Universit\"at, \\Theresienstra{\ss}e 37, 80333 M\"unchen, Germany\\
 \vskip 5pt
~$^3$Center for Particle Cosmology, Department of Physics and Astronomy, \\ University of Pennsylvania, 209 South 33rd St, Philadelphia, PA 19104, USA
 }

\end{center}  
 
\begin{abstract}

The idea of self-interacting bosonic dark matter capable of exhibiting superfluidity is revisited. We show that the most interesting parameter space of the theory corresponds to fully thermalized dark matter halos. As a result the entire halo undergoes Bose-Einstein condensation due to high degeneracy. Since it is observationally preferable for the dark matter density profile to be similar to cold dark matter in the outskirts of the halo, we argue that the Jeans wavelength must be at least few times shorter than the virial radius. This entails that, upon condensation, a dark matter halo fragments into superfluid clumps. However, we demonstrate that these would-be solitons experience strong tidal disruption and behave as virialized weakly interacting streams. An exception is the central soliton, which can be as large as few tens of kiloparsecs in size without contradicting observational bounds. As a result, in dwarf galaxies, the observed rotation curves can be completely contained within the superfluid soliton. In this case, the dark matter distribution is expected to be strongly sensitive to the baryonic density profile. We argue that the diversity of rotation curves observed for dwarf galaxies is a natural consequence of the superfluid dark matter scenario.

\end{abstract}

\newpage
\setcounter{page}{1}

\renewcommand{\thefootnote}{\arabic{footnote}}
\setcounter{footnote}{0}

\linespread{1.1}
\parskip 4pt

\section{Introduction}

The success of the Cold Dark Matter (CDM) paradigm is trumped only by its simplicity. In fact, it is incredible how such a simple model of the dark sector can provide an acceptable fit to various observations across the diverse range of scales, from galaxies to large-scale structures of the Universe. The simplicity of the model, in which one posits the existence of massive, purely gravitationally interacting particles may be interpreted as a statement both about the extent of observations and our ability to simulate the predictions of the model from first principles.

In fact, when it comes to understanding galactic-scale observations, one needs to tweak the computations by making additional assumptions about how baryonic processes affect the dark matter (DM) distribution. See~\cite{Bullock:2017xww,Salucci:2018hqu} for reviews of the small-scale shortcomings of CDM. At the end of the day, these puzzles boil down to the relative universality of the DM distribution in local structures, that stems from numerical simulations. The moment structure formation enters the nonlinear regime, the necessity of the numerical approach seems inevitable for making precise statements. 

In the absence of baryons, $N$-body simulations have established that CDM halos relax to a universal Navarro--Frank--White profile~\cite{Navarro:1995iw}. Baryons can be incorporated via hydrodynamic simulations ({\it e.g.},~\cite{Oman:2015xda}), and their complicated physics can play a major role in modifying the cuspy quasi-equilibrium CDM distribution to cored profiles preferred by observations. However, to fit observations, certain assumptions must be made about the stellar evolution history of galaxies. The extent to which such assumptions about feedback processes are justified is at the heart of the debate. The cuspy nature of CDM profiles also leads, among other things, to an over-prediction of dynamical friction in some systems~\cite{Debattista:1997bi,Sellwood:2016,Fornaxold,Tremaine}. 

There are also striking correlations between the dynamical acceleration of baryons and their distribution in disk galaxies, known as the radial acceleration relation~\cite{McGaugh:2016leg,Lelli:2016cui}. This is the generalization of the baryonic Tully-Fisher relation~\cite{McGaugh:2000sr,McGaugh:2011ac,Papastergis:2016jqv,LelliBTFR}, relating the fourth power of the asymptotic circular velocity to the total baryonic mass of disk galaxies. This is {\it a priori} surprising, given the intrinsically stochastic nature of baryonic feedback processes. Much progress has been made in recent years to understand the origin of such correlations in the context of CDM. Semi-empirical models with cored DM halos can reproduce the observed slope and normalization of such a relation, but not yet the small scatter~\cite{LelliBTFR,DiCintio:2015eeq,Desmond:2017}.

Alternatively, these potential shortcomings of CDM may be the indication that an extension of the model is warranted. Two main extensions have been followed. The first extension is to endow DM with self-interactions that are strong enough to alter the galactic core significantly~\cite{Spergel:1999mh}. See~\cite{Kaplinghat:2015aga} for the incorporation of self-interactions in DM simulations. A second extension is instead to ``fuzz out'' the cuspy profiles by considering the extremely light mass range ($m\lesssim 10^{-21}\,{\rm eV}$) for DM particles~\cite{Hu:2000ke}, hence the name Fuzzy Dark Matter. The idea is to consider particles that have de Broglie wavelength of order kpc, such that DM behaves as a Bose-Einstein condensate on astrophysical scales. Interested readers are referred to~\cite{Hui:2016ltb} for a detailed analysis of observational features of this model; see also~\cite{Ferreira:2020fam} for a recent review. Numerical simulations of this scenario, solving the Schr\"odinger--Poisson system of equations, have demonstrated the suppression of  substructures and formation of non-topological solitons in the core of the halos~\cite{May:2021wwp,Yavetz:2021pbc,Liu:2022rss}.

In this work we explore a third possible extension, which in some sense lies at the intersection of the two avenues mentioned above. Namely, we consider self-interacting sub-eV mass bosonic DM capable of superfluidity.
The idea of ameliorating the small-scale phenomenology of CDM with bosonic DM exhibiting superfluidity upon condensation goes back to~\cite{Goodman:2000tg}. It was originally suggested as an (improved) alternative to self-interacting DM (SIDM) model~\cite{Spergel:1999mh}, to suppress the formation of small-scale substructures. Over a certain mass range and scattering length, the core of bosonic DM halos can be supported at low densities, even at zero temperature. In contrast, in the original ($\sim {\rm GeV}$ mass) SIDM model, the cores could be prone to gravo-thermal collapse upon cooling. 

Later on, by combining galactic rotation curve observations and Bullet Cluster constraints~\cite{Markevitch:2003at,Clowe:2003tk}, it was argued that this original scenario of DM superfluidity was observationally ruled out~\cite{Slepian:2011ev}. A key ingredient in this argument was the finite-temperature equation of state for the theory in question. A more rigorous calculation of the equation of state in~\cite{Sharma:2018ydn,Sharma:2022jio} revealed some shortcomings in the original derivation. Furthermore, the original analysis of~\cite{Slepian:2011ev} relied on two important simplifying assumptions:~$i)$~global thermal equilibrium of the halo; and~$ii)$~spherical symmetry of the halo in Gross-Pitaevskii approximation. Relaxing the former assumption was originally explored in~\cite{Berezhiani:2017tth} in the context of an alternative (more exotic) scenario of DM superfluidity~\cite{Berezhiani:2015pia,Berezhiani:2015bqa}, aimed specifically at explaining the various galactic scaling relations. The consequences of relaxing both assumptions was explored in~\cite{Berezhiani:2021rjs}, in the context of the quartic model discussed in~\cite{Slepian:2011ev}, with alternative conclusions.

The goals of the current work is to further develop and understand the phenomenological implications of DM superfluidity. We will make the case for the existence of the parameter subspace for which galaxies are expected to host a superfluid core, with a coherence length exceeding tens of kiloparsecs. We will argue that this has important ramifications for reproducing the observed diversity of rotation curves of dwarf galaxies.

\section{Dark Matter Superfluidity}
\label{superfluid_intro}

The simplest realization of the scenario amounts to modeling DM as a massive scalar field with quartic potential,
\beq
\mathcal{L}=-\frac{1}{2}(\partial \phi)^2-\frac12 m^2\phi^2-\frac{\lambda}{4!}\phi^4\,.
\label{rel_lag}
\eeq
We assume that this bosonic sector is minimally-coupled to gravity, as usual. This DM model was independently studied in other contexts by~\cite{Peebles:1998qn,Peebles:1999fz}. 

Upon Bose-Einstein condensation this system exhibits superfluidity, provided that~$\lambda>0$. Consequently, at near-zero temperature the only permissible non-relativistic excitations are sound waves. Submerged non-relativistic objects can dissipate energy only through the production of these quanta, known as `phonons'. Therefore, for subsonic inertial motion, the substance appears to be frictionless, hence the name `superfluid'.

As argued in~\cite{Goodman:2000tg}, if DM described by~\eqref{rel_lag} reaches thermal equilibrium in galactic halos, then Bose-Einstein condensation can ensue for a certain particle mass range. The resulting substance is protected against gravitational collapse by means of the interaction pressure. The density profile of the self-gravitating DM core is easiest to find in the non-relativistic limit, wherein the exact spherically-symmetric density profile can be obtained analytically. At zero temperature, the superfluid equation of state is
\beq
P = \frac{\lambda \rho^2}{16m^4}\,.
\label{eos}
\eeq
The equations of hydrostatic equilibrium reduce to a Lane-Emden equation, with solution~\cite{Chandrasekhar} (see also~\cite{Goodman:2000tg})
\beq
\rho(r)=\rho_0 \frac{{\rm sin}(2\pi r/\ell)}{2\pi r/\ell}\,.
\label{soliton_profile}
\eeq
The central density~$\rho_0$ is an integration constant, which is fixed by the total mass of the core:
\beq
M_{\rm soliton} = \frac{\rho_0 \ell^3}{2\pi} \,.
\label{msoliton}
\eeq
The diameter~$\ell$ of the self-sustained configuration is given in terms of the parameters of the theory by
\beq
\ell\equiv\sqrt{\frac{\pi \lambda}{8Gm^4}}\,,
\label{ell def}
\eeq
with~$G$ denoting Newton's constant. As it will soon become apparent,~$\ell$ also coincides with the Jeans wavelength. Notice that for the quartic theory the soliton size is independent of the central density,
and therefore of the soliton mass. 

As long as we are only interested in the non-relativistic regime, the theory can be equivalently written in the Schr\"odinger form by taking the appropriate limit in which speed of light is taken to infinity. The resulting Lagrangian, including gravity, is
\beq
\mathcal{L}=\frac{1}{8\pi G}\Phi \vec{\nabla}^2 \Phi+\frac{{\rm i}}{2}\left( \psi^*\partial_t \psi-\psi\partial_t \psi^* \right)-\frac{|\vec{\nabla}\psi|^2}{2 m}-\frac{\lambda}{16m^2}|\psi|^4-m|\psi|^2\Phi\,,
\label{nonrel_lag}
\eeq
where~$\Phi$ is the Newtonian potential, while~$\psi$ is the complex field describing the dynamics of the original real bosonic degree of freedom~$\phi$. The equations of motion are Poisson's equation sourced by the DM density,~$\vec{\nabla}^2 \Phi = 4\pi G m|\psi|^2$, and a non-linear Schr\"odinger equation governing~$\psi$, known as the Gross-Pitaevskii equation.\footnote{The Schr\"odinger equation is as usual first-order in time, in contrast to the second-order equation for the original real field~$\phi$. This is precisely the reason the doubling of the field content.}

The non-relativistic formulation~\eqref{nonrel_lag} possesses a global~$U(1)$ symmetry, which is responsible for particle number conservation. However, no such symmetry and associated conservation law are present in the more 
fundamental relativistic description~\eqref{rel_lag}. For instance, as one can see from the latter, four~$\phi$ quanta can annihilate into two. Because such processes include relativistic particles, they are not accounted for by~\eqref{nonrel_lag}.

It is worthwhile mentioning that the theory considered in~\cite{Slepian:2011ev} was one for the relativistic complex scalar field instead of~\eqref{rel_lag}, reportedly due to the fact that the real scalar field theory would allow the depletion of the DM condensate via particle number changing processes discussed above. This is not of the central importance, since for galactic-scale analysis one works with a non-relativistic formulation anyway, and in this limit the difference between real and complex field scenarios disappears. Nevertheless, we would like to point out that although such a depletion of the galactic DM core indeed takes place, the rate for this process can be small. In particular, the time scale for significant depletion of the condensate via~$4\rightarrow 2$ annihilation channel can be estimated as~\cite{Dvali:2017eba,Dvali:2017ruz} (see also~\cite{Berezhiani:2021gph})
\beq
t_{\rm depl}\sim \left( \lambda m \left(\frac{\lambda n}{m^3} \right)^3 \right)^{-1}\,,
\label{depltime}
\eeq
where~$n$ is the number density of particles in the condensate. In the context of DM superfluidity in galaxies, the depletion time scale can be expressed in the following convenient form in units of the Hubble time:
\beq
\frac{t_{\rm depl}}{t_H}\sim 10^{33}\left( \frac{m}{\rm eV} \right)^{5} \left( \frac{\rho_{\rm DM}}{10^{-25}\, {\rm g}/{\rm cm}^3} \right)^{-3}\left(\frac{\sigma/m}{{\rm cm}^2/{\rm g}}\right)^{-2}\,.
\label{deplconst}
\eeq
Here, we have normalized the DM density~$\rho_{\rm DM}$ to its characteristic value~$10^{-25}\,{\rm g}/{\rm cm}^3$ reached in the galactic core. We also substituted the cross-section~$\sigma/m$ in terms of the parameters of the Lagrangian
\beq
\frac{{\sigma}}{m}=\frac{\lambda^2}{128\pi m^3}\,,
\label{sigoverm}
\eeq
and normalized to its value usually encountered in Bullet Cluster bounds.\footnote{For the purpose of deriving~\eqref{deplconst}, we only kept track of the parametric dependence~$\sigma/m \sim \lambda^2/m^3$, not of the numerical prefactor.} As one can see from~\eqref{deplconst}, the mass of the DM particle cannot be taken to values much lower than~$10^{-6}{\rm eV}$, as long as the interaction cross-section saturates the commonly interpreted Bullet Cluster bound~$\sigma/m<{\rm cm}^2/{\rm g}$. We will return in Sec.~\ref{bullet} to the discussion of the bound stemming from this depletion argument, where the parameter space will be analyzed in detail.

Be that as it may, as was correctly pointed out in~\cite{Slepian:2011ev}, this potential issue of depletion can be nullified by considering a fundamentally complex scalar field with a global charge. In this case, however, the production mechanism should be designed in such a way that the DM field configuration is produced in a state with large global charge. In particular, the charge density must be of order the particle density, which excludes the simplest vacuum misalignment mechanisms.

\section{Formation of Dark Matter Superfluid}
\label{formation DM}

In order to establish the qualitative picture of structure formation in the context of dark matter superfluidity, and the resulting density distributions in halos, we will assume that structure formation proceeds as in CDM until particles begin to scatter off each other. This is substantiated by numerical simulations for SIDM~\cite{Kaplinghat:2015aga}, albeit for the case of non-degenerate phase space for which DM is treated as a collisional classical gas.

In the CDM framework, DM particles are assumed to interact purely gravitationally, and the mass range is such that their de Broglie wavelength is much shorter than the inter-particle separation. Thus DM can be treated as a classical gas without interaction pressure. In this context, halo formation is a highly non-linear process, which requires numerical analysis. Ignoring the effect of baryons, DM-only ($N$-body) simulations reveal that the DM distribution in halos follows a universal Navarro-Frank-White (NFW) profile~\cite{Navarro:1995iw}
\begin{equation}
\rho(r)=\frac{\rho_{\rm NFW}}{\frac{r}{r_s}\left(1+\frac{r}{r_s}\right)^2}\,.
\label{eq:NFW}
\end{equation}
The characteristic density~$\rho_{\rm NFW}$ and scale parameter~$r_s$ are fitting parameters that vary from halo to halo. The size of the halo itself is typically defined in terms of the virial radius,~$R_{\rm V}$, the radius within which the mean DM density is~200 times the critical density~$\rho_{\rm c}$. The ratio between the virial radius and the scale parameter defines the concentration parameter,~$c=R_{\rm V}/r_s$, which is connected to the total mass by the mass-to-concentration relation~\cite{Dutton:2014xda}. For reference, typical parameter values for a Milky Way-like galaxy are~$\rho_{\rm NFW} =  10^{-25} \text{g}/\text{cm}^{3}$,~$c= 6$ and~$R_{\rm V} = 200$~kpc. 

In the galactic environment, the superfluid phase transition may be achieved by Bose-Einstein condensation. From an effective field theory point of view, the condensate is described by a classical field configuration with finite number density, which spontaneously breaks the~$U(1)$ global symmetry of the underlying Lagrangian. At sufficiently low momentum/energy, the dynamics of the scalar field are described by phonons, {\it i.e.}, gapless perturbations of the classical configuration. The spectrum of fluctuations around the homogeneous condensate is given by ({\it e.g.},~\cite{Berezhiani:2019pzd})
\begin{equation}
\omega_k^2=-4\pi G\rho+c_s^2k^2+\frac{k^4}{4m^2}\,, 
\label{eq:dispersion} 
\end{equation}  
where~$\rho$ is the superfluid density, and
\beq
c_s^2 =\frac{\partial P}{\partial \rho} =\frac{\lambda \rho}{8m^4}
\eeq
is the adiabatic sound speed. The first term in~\eqref{eq:dispersion} is the tachyonic contribution associated with the gravitational instability of the system; the second term represents the energy cost to excite a sound wave of momentum~$k$; and the third term is the kinetic energy of a constituent of the system.

\subsection{Jeans scale}

From the dispersion relation~\eqref{eq:dispersion}, it is easy to see that modes softer than the scale~$k_{\rm J}$ are unstable, with
\begin{equation}
k_{\rm J}^2=2 m^2 c_s^2\left(-1+\sqrt{1+\frac{4\pi G \rho}{m^2 c_s^4}}\right)\,.
\label{eq:JS}
\end{equation} 
The corresponding Jeans length scale, ~$\ell=2\pi/k_{\rm J}$,  sets an upper bound on the size of a superfluid phase of the DM halo which is gravitationally stable~\cite{Berezhiani:2021rjs}.

Evidently, the Jeans momentum~\eqref{eq:JS} depends on the relative strength of self-interactions and gravity. It is useful to consider two different regimes, depending on the magnitude of the following dimensionless parameter:
\begin{equation}
\xi=\frac{m^2 c_s^4}{4\pi G\rho}\,.
\label{eq:xi}
\end{equation}
This parameter, which appears under the square root in~\eqref{eq:JS}, encodes the nature of the pressure sustaining the core. 

If self-interactions are too feeble, such that~$\xi \ll 1$, then gravitational collapse is prevented by the ``quantum pressure'' of the system. The Jeans scale in this case reduces to 
\begin{equation}
\ell\simeq \left({\frac{\pi^3}{ G \rho m^2}}\right)^{1/4} \qquad (\xi \ll 1)\,.
\end{equation}
This regime is relevant Fuzzy Dark Matter models. In the opposite regime~$\xi \gg 1$, the superfluid core is sustained by the pressure due to self-interactions. The Jeans scale in this case reduces to
\begin{equation}
\ell\simeq \sqrt{\frac{\pi c_s^2}{ G\rho}} \qquad (\xi \gg 1)\,.
\label{eq:interacting}
\end{equation}
This is the regime of interest for our purposes, and we will henceforth refer to it as the \textit{interaction pressure case}.

Because of this instability, the would-be homogeneous condensate in the absence of gravity, will in general break up into a cluster of gravitationally-stable lumps. The detailed substructure of these lumps depends on the nature of the fluctuations seeding the instability. In case of {\it spherically-symmetric} Jeans collapse, as assumed in~\cite{Slepian:2011ev,Sharma:2018ydn}, the thermalized region would form a single solitonic superfluid core~\eqref{soliton_profile}, surrounded by an envelope of weakly-interacting particles supported mainly by statistical pressure ({\it i.e.}, virial motion). Assuming that the collapse is adiabatic, maintaining equilibrium even in the envelope, the result would be an 
isothermal density profile in the envelope, with a core-envelope matching condition determined by the finite-temperature equation of state of the system. It is this particular scenario (spherically-symmetric, adiabatic collapse) which was argued to be ruled out in~\cite{Slepian:2011ev}. If the collapse is out-of-equilibrium, on the other hand, then the envelope would need to rethermalize in order achieve isothermality. Otherwise, it would be more natural to expect an NFW profile in the envelope, as assumed in~\cite{Berezhiani:2017tth}.

Importantly, it was argued in~\cite{Berezhiani:2021rjs} that the assumption of spherical symmetry seems hard to justify, and one should consider more general destabilizing fluctuations. 
In the more general case, the resulting fragmented thermal region will be comprised of a collection of solitons, whose individual density profiles are given by~\eqref{soliton_profile}, together with
a significant fraction of left-over particles in the disordered phase. Furthermore, this was shown to entail severe Bullet Cluster constraints on the parameter space, such that the superfluid solitons must 
be smaller than a few kpcs in diameter. 

The goal of the present work is to revisit this scenario and further explore its observational consequences. In particular, we will discuss additional effects that have important ramifications for halo formation and the resulting superfluid density profiles. 

\subsection{Size of thermalized region}
\label{thermal size}

As discussed above, the rearrangement of the phase space, relative to CDM, is not expected to happen everywhere throughout the halo. Bose-Einstein condensation requires bosons to reach the maximal entropy state, which is achieved via interactions. This is also referred to as reaching equilibrium, or thermalization. Even if DM were already thermalized by the time structures form, one expects that violent relaxation inherent to virialization would most likely 
kick the system out of equilibrium. Thus, particles must re-thermalize within local structures in order to Bose-Einstein condense. Since high densities facilitate equilibration by enhancing scattering rates, thermalization is expected to be more efficient in the core of the halo than in the outskirts.

Let us denote by~$R_T$ the radius within which particles have interacted enough to reach thermal equilibrium. We will henceforth refer to~$R_T$ as the `thermal radius'. It can be simply estimated by demanding that particles within~$R_T$ have experienced at least one interaction over the lifetime of the halo~\cite{Sikivie:2009qn,Erken:2011dz}. This oft-made approximation is of course quite crude, since more than a single scattering should be necessary to achieve thermal equilibrium. In any case, the condition is
\begin{equation}
\Gamma(R_T) t_g=1\,,
\label{eq:conditionRT}
\end{equation}
where~$t_g\sim 10$ Gyr is the typical lifetime of a galaxy, and~$\Gamma$ is the relaxation rate. For the theory at hand, the latter can be approximated~$\Gamma$ as the 2-body scattering rate:
 \begin{equation}
\Gamma =(1+\mathcal{N})\frac{\sigma}{m}\rho v\,.
 \label{eq:csRate}
\end{equation}
Importantly, we see that the relaxation rate depends on the DM density profile~$\rho(r)$ of the galaxy at hand. It also depends on the one-dimensional velocity dispersion~$v(r)$ of DM particles before the phase transition. For simplicity, we will assume that~$v(r)$ is of the same order as the mean orbital velocity at~$r$. The relaxation rate also depends on the parameters of the theory through the scattering cross section~\eqref{sigoverm}.  
Lastly,~$\mathcal{N}$ is the Bose-enhancement factor, 
\beq
\mathcal{N}=\frac{\rho}{m}\left(\frac{2\pi}{m v}\right)^3  \,,
\eeq
which accounts for the fact that particles are scattering into a highly degenerate phase space. Clearly the Bose-enhancement factor entails further dependence on~$\rho(r)$.

\subsection{Superfluid subhalos from fragmentation}
\label{thermal fragment}

The final density distribution of an inner galactic region depends sensitively on the relative magnitude of~$\ell$ and~$R_T$. Specifically,

\begin{itemize}
\item If~$\ell> R_T$, then a gravitationally-stable superfluid core of size~$R_T$ will be formed.

\item If~$\ell<R_T$, then the Jeans instability will be triggered inside the thermalized core, resulting in fragmentation.

\end{itemize}

\noindent As discussed in~\cite{Berezhiani:2021rjs}, only the latter hierarchy,~$\ell\lesssim R_T $, is phenomenologically viable in the interaction pressure case~$\xi\gg 1$ of interest. {\it This implies that the thermal core is unstable, and is prone to fragmenting into a collection of superfluid droplets of size~$\ell$.} 

The density profile within each droplet would approximatively be given by the density distribution~\eqref{soliton_profile} of a self-gravitating zero temperature configuration. 
In other words, such a destabilization would, in general, be expected to form a cluster of superfluid solitons, similar to the formation of star clusters from the large enough
clouds. As we will see in Sec.~\ref{tidal}, however, the superfluid solitons formed through fragmentations are susceptible to tidal disruption, which happens on the
same time scale as fragmentation. 

\subsection{Superfluid subhalos from hierarchical structure formation}
\label{pre-existing subhalos}

On top of the solitons formed through the fragmentation instability discussed above, a galactic halo is also host to a population of subhalos, due to the well-known hierarchical nature of structure formation.
In numerical simulations for CDM and SIDM~\cite{Vogelsberger:2012ku,Robles:2019mfq}, the mass function of resolved subhalos has been shown to be quite similar in both models, over the
range~$4.5\;{\rm km}/{\rm s} \lesssim V_{\rm max} \lesssim 40\;{\rm km}/{\rm s}$, where~$V_{\rm max}$ is the maximum subhalo circular velocity. The corresponding resolved mass range is approximately
\beq
5 \times 10^6\;M_\odot \lesssim M \lesssim 5\times10^7 M_\odot \,.
\eeq

Our working assumption is that structure formation proceeds in the early stages as in CDM, before thermalization rearranges the phase space distribution significantly.
Hence a similar population of subhalos should be present in the superfluid scenario, with one important caveat. Once a given subhalo thermalizes and undergoes the
superfluid phase transition, we expect its inner region to ``fuzz out'' and relax to the cored profile~\eqref{soliton_profile} up to size~$\ell$, thereby lowering its central density. 

As is the case in ultra-light DM models~\cite{Hui:2016ltb,Berezhiani:2017tth}, consistency requires that the resulting central density~$\rho_0$ be greater
than the virial density~$\rho_{200} = 200 \frac{3H_0^2}{8\pi G} \simeq 1.95\times 10^{-27}\;{\rm g}/{\rm cm}^3$. This translates into a lower bound on the
halo mass of
\beq
M_{\rm min} \simeq 5\times 10^3  \left(\frac{\ell}{\rm kpc}\right)^3 M_\odot\,.
\label{Mmin}
\eeq
Therefore, the CDM mass function should be truncated at~$M_{\rm min}$. While this may seem innocuous for~$\ell\simeq {\rm kpc}$, notice that~$M_{\rm min}$ becomes comparable
to the mass of the lightest resolved subhalo already for~$\ell\simeq 10~{\rm kpc}$.

\section{Tidal Disruption of Solitons}
\label{tidal}

An important effect, not discussed in~\cite{Berezhiani:2021rjs}, concerns the tidal disruption of the superfluid solitons. Both the Jeans scale~\eqref{eq:JS} and the density profile~\eqref{soliton_profile} were derived assuming that the superfluid phase is not affected by any other external potential. While this may be justified for solitons near the center of the halo, superfluid droplets far enough in the outskirts will experience the gravitational potential sourced by other solitons within their orbit. We first estimate the impact of tidal disruption in generality, and then discuss its implications in turn for superfluid solitons generated through fragmentation, and those generated through hierarchical structure formation. 

\subsection{Tidal radius}

To quantify the strength of tidal effects, it is useful to introduce the tidal radius~$r_\text{tidal}$~\cite{Kesden:2006vz}:
\begin{equation}
 r_\text{tidal}=r\left(\frac{M_\text{soliton}}{M(r)}\right)^{1/3}\,.
\end{equation}
This quantifies the stability of a soliton of mass~$M_{\rm soliton}$, orbiting at a distance~$r$, due to the presence of an enclosed mass distribution~$M(r)$. Tidal disruption of the soliton takes place if its tidal radius is smaller than its size, {\it i.e.}, if~$r_{\rm tidal} < \ell$. In this case, tidal forces are strong enough to dominate over internal binding forces, and the superfluid droplet shatters, leaving behind distorted fluid debris in the form of streams. This can also be understood from the point of view of the phonon spectrum: For~$r_\text{tidal}<\ell$ the dispersion relation~\eqref{eq:dispersion} will receive important anisotropic corrections from the external gravitational potential, and consequently the coherence length is expected to shrink in some directions and stretch in others.

To compare~$r_\text{tidal}$ and~$\ell$, we must estimate the density distribution generated by a collection of superfluid solitons sitting within an orbit of radius~$r$. For this purpose, we will assume this distribution to be given by the NFW profile~\eqref{eq:NFW}. This can be justified as follows. Although we have simplistically described the formation of superfluid droplets as a sequence of three events --- thermalization, phase transition, and fragmentation --- these are expected to happen in conjunction. Small fractions of high-density regions of the halo are expected to thermalize, become superfluid and merge until a soliton of size~$\ell$ is formed. Thermalization keeps on affecting regions of lower density, and again solitons are formed there. Therefore, we expect that the average density of each soliton should be connected to the local density of the halo in which they formed. Because the NFW profile was our starting input, the density distribution of droplets within a radius~$r$ is expected to follow an NFW profile as well.

Using~\eqref{eq:NFW} to compute the mass distribution~$M(r)$ of an NFW halo, together with~\eqref{msoliton} for~$m_\text{soliton}$, we obtain:
\begin{equation}
\frac{r_\text{tidal}}{\ell}\simeq \begin{cases}\left(\frac{\rho_0}{40 \rho_\text{NFW}}\right)^{1/3} \left(\frac{r}{r_s}\right)^{1/3} & \mbox{for}~r\ll r_s\,; \\ \left(\frac{\rho_0}{80 \rho_{\text{NFW}}}\right)^{1/3}\frac{r}{r_s} & \mbox{for}~r\gg  r _s\,,\end{cases}
\label{eq:tidal2}
\end{equation}
where~$\rho_{\rm NFW}$ is the characteristic density of the NFW profile. Since the NFW profile~\eqref{eq:NFW} itself behaves as~$\rho(r) \simeq \rho_\text{NFW} \frac{r_s}{r}$ for~$r \ll r_s$, and~$\rho(r) \simeq \rho_\text{NFW} \left(\frac{r_s}{r}\right)^3$ for~$r \gg r_s$, the above criterion can be summed up simply as
\be
\frac{r_\text{tidal}}{\ell}\simeq \left(\frac{\rho_0}{40 \rho(r)}\right)^{1/3}\,.
\ee
Therefore, solitons at a distance~$r$ from the center will be immune to tidal disruption if their core density is at least 40 times the average DM density in their vicinity, {\it i.e.},~$\rho_0 \gtrsim 40 \rho(r)$. 
However, superfluid solitons do not follow circular orbits in general; for instance, those on eccentric orbits will go through the inner region where the local density is high. A more realistic, and generally applicable
lower bound is obtained by replacing the local density~$\rho(r)$ with the characteristic NFW density~$\rho_{\rm NFW}$:
\be
\rho_0 \gtrsim 40 \rho_{\rm NFW} \qquad \text{(no tidal disruption)}\,.
\label{rho0 bound}
\ee
We next study the implications of this bound in turn for solitons that are formed through the fragmentation instability, and those that are formed from pre-existing subhalos. 

\subsection{Tidal disruption of subhalos generated by fragmentation}

For solitons that formed via fragmentation, it is instructive to translate~\eqref{rho0 bound} into a lower bound on the size~$R$ of the initial seed fluctuation that gave rise to the soliton. Equating the mass~$M_{\rm fluc} \simeq \rho(r) R^3$ of an initial (not necessarily spherical) fluctuation, to the mass~$m_\text{soliton}=\frac{\rho_0 \ell^3}{2\pi}$ of the eventual soliton, we obtain
\beq
R = \left(\frac{1}{2\pi} \frac{\rho_0}{\rho(r)}\right)^{1/3} \ell \gtrsim 2\,\ell\,,
\label{Rbound}
\eeq
where in the last step we have used~\eqref{rho0 bound}. Thus it seems that sufficiently large initial fluctuations would result in tidally-stable solitons. 

However, one should bear in mind that this bound is oversimplified, since we are generally interested in~$\ell\gtrsim {\rm kpc}$, and densities in the assumed set-up may not be homogeneous on such scales. Furthermore, before an overdensity significantly larger than~$\ell$ has the opportunity to collapse into a dense soliton, it will already be susceptible to tidal disruption. In other words, tidal disruption may prevent the soliton from forming in the first place. Fragmentation and tidal disruption take place in parallel, and on the same time scale of order the dynamical time~$t_{\rm dyn} \simeq 1/\sqrt{G\rho}$. Therefore, we expect that few (if any) of the solitons generated through fragmentation will survive tidal disruption.

\subsection{Tidal disruption of pre-existing substructure}

Next let us consider the impact of tidal disruption on superfluid solitons formed as subhalos in hierarchical structure formation. As argued in Sec.~\ref{pre-existing subhalos}, this population of solitons should approximately follow the subhalo mass function of CDM, with a low-mass cutoff of~$M_{\rm min} = 5\times 10^3  \left(\ell/{\rm kpc}\right)^3 M_\odot$.

The lower bound~\eqref{rho0 bound} on the central density, which ensures robustness against tidal disruption, translates to a lower bound on the soliton mass via~\eqref{msoliton}:~$M \gtrsim 40 \frac{\rho_{\rm NFW}} {2\pi} \ell^3$.
Substituting the characteristic density~$\rho_{\rm NFW} =  10^{-25} \text{g}/\text{cm}^{3}$ for a Milky Way-like host halo, this gives
\beq
M \gtrsim 10^7 \left(\frac{\ell}{{\rm kpc}}\right)^3 M_\odot\,.
\eeq
When compared to the CDM subhalo mass function, we see that this lower bound is comparable to the heaviest subhalos found in the simulations for~$\ell \gtrsim \text{few kpc}$. 
As our interest lies with the parameter space for which the Jeans scale significantly exceeds a few kpcs, we can safely conclude that none of the subhalos generated through hierarchical
structure formation would survive tidal effects in our context. 

Obviously, resolved subhalos amount to a small fraction of the total mass of a CDM halo. The rest of the DM is in the form of virialized dust. In our case, the phase space for some of this dust would be rearranged by thermalization. In light of our discussion of fragmentation and tidal disruption above, this thermalized component would come in the form of a turbulent fluid, similar in this sense to its Fuzzy DM counterpart~\cite{Liu:2022rss}, including streams of superfluid debris. Another interesting recent work \cite{Glennon:2022huu} studied the impact of self-interactions on the tidal disruption of Fuzzy DM solitons orbiting the host NFW halo.

\subsection{Overall Structure}

Combining the insights derived in Secs.~\ref{formation DM} and~\ref{tidal} thus far, we arrive at the following overall picture of the superfluid DM distribution in galactic halos.  

\begin{itemize}

\item At distances~$r>R_T$ from the center, particles are expected to be oblivious to the contact interactions and therefore follow a velocity distribution similar to CDM, albeit with highly degenerate phase space (due to the sub-eV mass range). In other words, the de Broglie wavelength of such particles will be significantly longer than the characteristic inter-particle separation.

\item There will be a central superfluid soliton of size~$\ell$. Its unavoidable formation stems from the fact that thermalization, and consequently the superfluid phase transition, begin from the highest density region, {\it i.e.}, the central region. This soliton is not expected to be tidally disrupted due to its special location. Assuming a central dark density of order~$10^{-25}{\rm g}/{\rm cm}^3$, we arrive at the following estimate for the mass of this central soliton
\beq
M_{\rm soliton} \simeq 2\times 10^{5}  \left(\frac{\ell}{\rm kpc}\right)^3 M_\odot\,.
\eeq

\item For~$\ell<R_T$, which as described in Sec.~\ref{thermal fragment} is the regime of interest, the DM enclosed within~$r<R_T$, except for the central soliton, will present itself in the form of virialized superfluid streams. These streams are the product of the fragmentation and tidal disruption instabilities discussed above. One can give a back-of-the-envelope estimate the velocity of the streams as follows. The energy imparted by tidal disruption on soliton at a distance~$r$ is
\beq
E_\text{tidal}=\left(\frac{M_{\rm soliton}}{M(r)}\right)^{2/3} \frac{GM(r) M_{\rm soliton}}{r}\,.
\eeq
Estimating the enclosed mass~$M(r)$ using the NFW profile~\eqref{eq:NFW}, the soliton mass as~$M_{\rm soliton} \sim \rho(r) \ell^3$, and using~\eqref{ell def}, it is straightforward to obtain
\beq
E_\text{tidal} \sim M_{\rm soliton} \rho \frac{\lambda}{8 m^4} \sim M_{\rm soliton}c_s^2\,.
\eeq  
Therefore, interestingly, the DM substructure streams are expected to be nearly sonic. 

\item The coarse-grained density profile outside of the central soliton, obtained by averaging over timescales longer than the dynamical time, is expected to have features of the NFW profile, stemming from the fact that both superfluid debris and the gaseous DM in the envelope are stabilized by virialization.

\end{itemize}

It will be interesting to verify whether this overall qualitative picture is borne out by numerical simulations of DM superfluidity.

\section{Revisiting the Bullet Cluster}
\label{bullet}

A well-known constraint on DM self-interactions comes from the analysis of the Bullet Cluster event~\cite{Markevitch:2003at,Clowe:2003tk}. This is the aftermath of two colliding clusters, the ``bullet'' and the target cluster, in which the gas component is observed to be offset with respect to the DM component. Simulations can accurately reproduce the event by assuming that the DM component behaves as a collisionless, purely gravitationally-interacting fluid during the collision.

In other words, the mean number of scatterings that a DM particle from the Bullet Cluster undergoes while passing through the target cluster must be negligible~\cite{Markevitch:2003at}:
\begin{equation}
\langle n_\text{sc}\rangle<1\,.
\end{equation}
The mean number of scatterings,~$\langle n_{sc}\rangle~$, is averaged over the entire Bullet Cluster volume, and can be estimated in terms of the interaction rate as:
\begin{equation}
\langle n_{sc}\rangle= \Gamma \frac{2 R_{\rm V}}{v_\text{infall}}\,,
\label{eq:RateBull}
\end{equation}
where~$v_\text{infall}\simeq10^{-2}$ is the in-fall velocity, and~$R_{\rm V}\simeq 2 \,\text{Mpc}$ is the virial radius of the target cluster. 
Assuming that DM is non-degenerate and interacts primarily via 2-body interactions, the constraint on~$\Gamma$ translates to 
\begin{equation}
\frac{\sigma}{m}\lesssim \frac{\text{cm}^2}{\text{g}}\,,
\label{eq:bulletnondeg}
\end{equation}
This is the well-known bound on DM self-interactions, widely employed in the literature.
It is important to stress that the sole general constraint we may infer from the Bullet Cluster is on the magnitude of the interaction rate, while any further statements are model dependent.
For instance, the constraint~\eqref{eq:bulletnondeg} is valid under the assumption of non-degeneracy and 2-body scattering.

It was argued in~\cite{Berezhiani:2021rjs} that if the DM phase space is degenerate in clusters, then the bound on~$\sigma/m$ must be revised to incorporate the Bose enhancement factor~${\cal N}$ in the interaction rate. 
Using the rate~\eqref{eq:csRate} appropriate for degenerate DM, the following tighter bound is obtained~\cite{Berezhiani:2021rjs}:
\begin{equation}
\frac{\sigma}{m}<\frac{1}{\mathcal{N}} \frac{\text{cm}^2}{\text{g}}\simeq 10^{-2} \left(\frac{m}{\text{eV}}\right)^4\frac{\text{cm}^2}{\text{g}}\,.
\label{eq:deg}
\end{equation}
The pillar of the argument is the assumption that a significant fraction of the halo is in the gaseous form, with velocity distribution similar to that of an NFW profile. As a result, for sub-eV mass range, DM interactions receive the 
aforementioned Bose enhancement. In other words, the same enhancement factor that makes DM thermalization possible over large galactic regions, also leads in general to a tighter Bullet Cluster bound.

A loophole we would like to pursue here, however, concerns the case of complete thermalization of halos. Whether the constraint~\eqref{eq:deg} can be circumvented in this way is contingent on the left-over gaseous DM as a result of the fragmentation of the entire halo into superfluid islands. This idea was briefly entertained in~\cite{Berezhiani:2021rjs}, however, as tidal disruption was not taken into account, the idea of having the outskirts of halos predominantly in the form of the solitonic substructure did not seem inviting.

Therefore, in the parameter subspace where interactions are strong enough for halos to experience thermalization and the subsequent phase transition into the tidally-disrupted fluid form, the relaxation of the Bullet Cluster bound relies on the absence of a significant gaseous component. At the level of our consideration this is impossible to evaluate rigorously, thus we will simply analyze the consequences of removing the enhancement factor from the merger analysis while at the same time keeping it in the thermalization computation.

As argued in Sec.~\ref{tidal}, the coarse-grained density profile of the turbulent fluid is expected to follow an NFW profile. Let us therefore take the density distribution in the target cluster to be NFW, with concentration parameter~$c\simeq 4$, characteristic density~$\rho_\text{NFW}\simeq 10^{-25} \text{g}/\text{cm}^3$, and virial radius~$R_{\rm V}=2 \,{\rm Mpc}$. With these values, it is straightforward to show that, if the cross-section is large enough to satisfy
\beq
\frac{\sigma}{m} \gtrsim 3\left(\frac{m}{\text{eV}}\right)^4\frac{{\rm cm}^2}{\rm g} \,,
\label{tmp_cross}
\eeq
then particles would have interacted sufficiently for halos to reach equilibrium in their entirety. In this case, the DM phase transition from gaseous to fluid form would take place, and the cluster merger would proceed without the Bose enhancement factor, per our argument given above. In this case, the Bullet Cluster constraint reduces to the familiar one, {\it i.e.},~$\frac{\sigma}{m} \lesssim \frac{\text{cm}^2}{\text{g}}$. Therefore, in this case of complete thermality, the allowed range of cross section becomes
\beq
3\left(\frac{m}{\text{eV}}\right)^4\frac{{\rm cm}^2}{\rm g} \lesssim \frac{\sigma}{m}\lesssim\frac{{\rm cm}^2}{\rm g} \qquad \text{(allowed; complete thermality)}\,.
\label{allowed complete}
\eeq

By contrast, if~\eqref{tmp_cross} does not hold, the presence of a significant gaseous component in clusters is unavoidable, and DM interactions in the merger will be enhanced by the Bose factor. 
In this case, the tighter Bullet constraint~\eqref{eq:deg} for degenerate DM applies:
\beq
\frac{\sigma}{m} \lesssim 10^{-2} \left(\frac{m}{\text{eV}}\right)^4\frac{\text{cm}^2}{\text{g}}     \qquad \text{(allowed; partial thermality)}\,, 
\label{allowed partial}
\eeq
This was the case considered in~\cite{Berezhiani:2021rjs}. 

Combining~\eqref{allowed complete} and~\eqref{allowed partial}, we conclude that the range of cross section
\begin{equation}
10^{-2} \left(\frac{m}{\text{eV}}\right)^4\frac{{\rm cm}^2}{\rm g}\lesssim\frac{\sigma}{m}\lesssim3\left(\frac{m}{\text{eV}}\right)^4\frac{{\rm cm}^2}{\rm g} 
\label{eq:exclband} 
\end{equation}
is excluded for sub-eV particles. Of course, the regime of very large cross-section,~$\frac{\sigma}{m}\gsim \frac{\text{cm}^2}{\text{g}}$, is excluded as well.

\begin{figure} 
	\centering
	\includegraphics[scale=0.45]{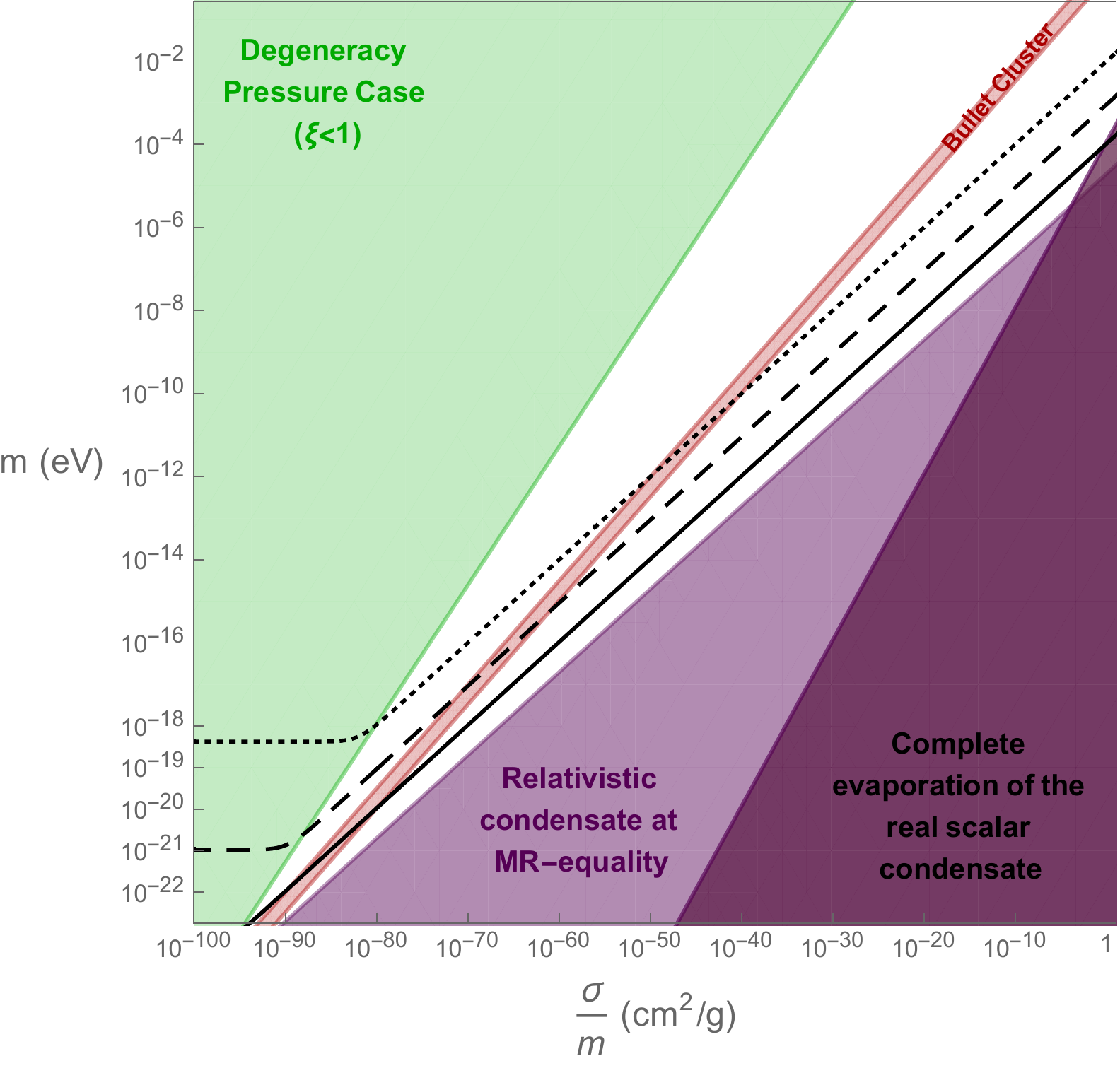} 	
\caption{\small{The parameter space for superfluid DM with quartic potential. The {\it Light purple} region is excluded by the requirement that the DM condensate is non-relativistic at matter-radiation equality, while the red stripe by the improved bullet cluster~\eqref{eq:exclband}. The {\it Green} region represents condensates which are dominated by degeneracy pressure, with negligible interaction pressure. (This is the regime relevant for Fuzzy DM.) The {\it Dark purple} region is only relevant for the real scalar field. In this region, a significant fraction of the DM is expected to evaporate into relativistic particles at matter-radiation equality. The dotted, dashed and solid lines correspond to Jeans length of~0.1,~2, and~30~kpc, respectively. As we can see, the 30 kpc line lies in the region where clusters and galaxies are in global thermal equilibrium.}}
	\label{fig:Fig1}
\end{figure}

We report in Fig.~\ref{fig:Fig1} the final parameter space of the DM superfluid candidate. 

\begin{itemize}

\item The green region corresponds to DM condensates that are dominated by quantum degeneracy pressure~($\xi \ll 1$), with negligible interaction pressure. This ``fuzzy DM'' regime is not ruled out per se, but
is not relevant for our scenario.  

\item The red stripe represents the excluded range~\eqref{eq:exclband} for degenerate DM.

\item The light purple region is excluded, as the condensate would have a relativistic equation of state at matter-radiation equality. 

\item The dark purple region is the result of invoking the depletion argument~\eqref{depltime} at the time of matter-radiation equality. This is relevant only if the underlying fundamental description of the DM particles is in the form of a real scalar field. In this parameter subspace, the significant fraction of the DM condensate is expected to evaporate into relativistic quanta within a Hubble time (at equality). Notice that the most of this region is already excluded by the light purple region (relativistic equation of state). However, in the higher mass-range corner, the depletion consideration yields an additional constraint.

\end{itemize}

\noindent We can separate two scenarios depending on what part of the allowed (white) region of the plot we are considering:

\begin{itemize} 
\item The white region to the left of the red stripe represents galaxies and clusters which are only locally in thermal equilibrium ($R_T<R_{\rm V}$). In this regime, interaction pressure can account for solitons with a size of at most~$\ell \simeq 6$~kpc.

\item The white region to the right of the red stripe represents galaxies and clusters which are globally thermalized. In this case, the size of the core can reach~$\ell \sim {\cal O}(10)~{\rm kpc}$, before spoiling predictions from galactic dynamics.

\end{itemize}

The dotted, dashed and solid lines correspond to Jeans length of~0.1,~2, and~30~kpc, respectively. To see how one can comfortably choose parameters for which the superfluid soliton is tens of kpcs in size, it is useful to express the equation of state at matter-radiation equality in terms of the Jeans scale,
\beq
\left.\frac{P}{\rho}\right\vert_{\rm equality}\simeq 10^{-5} \left( \frac{\ell}{\rm kpc} \right)^2\,.
\eeq
This is obtained by combining~\eqref{eos} and~\eqref{ell def}, and substituting the density at equality,~$\rho_{\rm equality} \simeq  4\times 10^{-20} {\rm g}/{\rm cm}^3$. Notice that, interestingly, the equation of state depends on the DM particle mass and cross section through the combination that determines the soliton diameter~$\ell$ as well. For~$\ell=100~{\rm kpc}$, we have~$P/\rho|_{\rm equality}\simeq 0.1$, which is borderline consistent with observations. In contrast, a diameter of~$\ell=30~{\rm kpc}$ gives~$P/\rho|_{\rm equality}\simeq 0.01$, which satisfies observational constraints.

Let us conclude this Section by emphasizing that our analysis leading to Fig.~\ref{fig:Fig1} has been based on estimates rather than a detailed computation. A more rigorous analysis of thermalization in our scenario, combined with numerical simulations of the Bullet cluster merger, would refine the allowed/excluded parameter regions. For instance, the excluded parameter strip and purple regions could widen or even shrink. Such a detailed numerical analysis is beyond the scope of this paper, and is left for future work.

\section{Rotation Curves}

The possibility of having large central solitons has interesting ramifications for galactic rotation curves. As mentioned earlier, an intriguing implication of the quartic theory is that the soliton size~$\ell$ is independent of the 
central density (or, alternatively, of the soliton mass). Thus the size of the central soliton will be the same in all galaxies. 

A convenient expression for~$\ell$ follows from~\eqref{ell def} and~\eqref{sigoverm}:
\beq
\ell=2 \left( \frac{\sigma/m}{{\rm cm}^2/{\rm g}} \right)^{1/4}\left( \frac{m}{\rm meV} \right)^{-5/4}\,{\rm kpc}\,.
\eeq
With~$\ell\simeq 30~{\rm kpc}$, for instance, the rotation curve of Milky Way-like galaxies will be mostly driven by the virialized superfluid debris, with the exception of the central core where baryons dominate the dynamics anyway. 
In dwarf galaxies, on the other hand, the entire observed rotation curves will be engulfed by the central superfluid soliton, as rotation curves tend to extend up to~$10$--$15\,{\rm kpc}$ from the center. Interestingly, these are precisely the types of galaxies that are hardest to fit within CDM~\cite{Oman:2015xda}, requiring the significant baryonic feedback even for low surface brightness (LSB) galaxies. 

In the context of the superfluid DM scenario, the situation depends on the distribution of baryons. To illustrate this, following~\cite{Berezhiani:2017tth} we consider a toy baryonic distribution, 
with two-parameter ``spherical exponential'' profile
\beq
\rho_\text{b}(r)=\frac{\gamma}{8\pi L^3}e^{- r/L}\,,
\label{rhob}
\eeq
where~$L$ plays the role of a radial ``scale length''. In the upcoming figures, we consider~$L=1$~kpc (orange lines) and~$L=10$~kpc (red lines), modeling respectively cusped and quasi-homogeneous
baryonic content. The central DM density and normalization constant~$\gamma$ are chosen so that the enclosed mass of DM and baryons in the soliton is~$M_{\rm DM}=10^{10}\,M_\odot$ and~$M_{\rm b}=2\times 10^9\,M_\odot$. Meanwhile, the theory parameters~$m$ and~$\lambda$ are fixed to achieve a soliton diameter of~$\ell=30$~kpc in the absence of baryons (dashed grey lines).

\begin{figure} 
	\centering
	\includegraphics[scale=0.5]{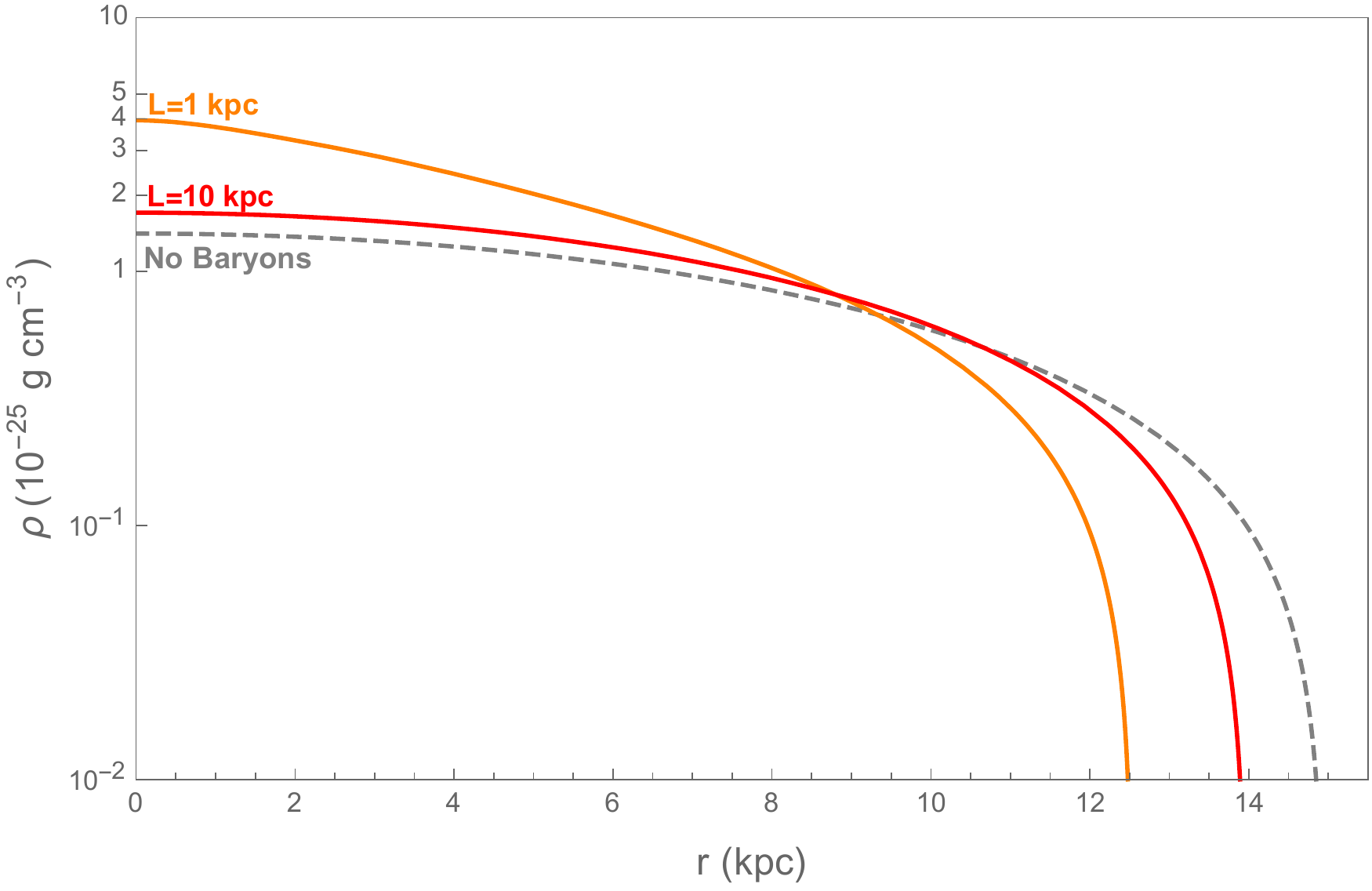} 	
  \caption{{\small{DM density distribution without baryon (dashed grey), and with a baryonic spherical exponential component~\eqref{rhob} with~$L=1$~kpc (orange) and~$L=10$~kpc (red).
  This shows how the DM soliton gets smaller for cuspy baryonic distributions, and the central dark matter density increases accordingly.}}}
\label{DM density prof}
\end{figure}

\begin{figure} 
	\centering
	\includegraphics[scale=0.5]{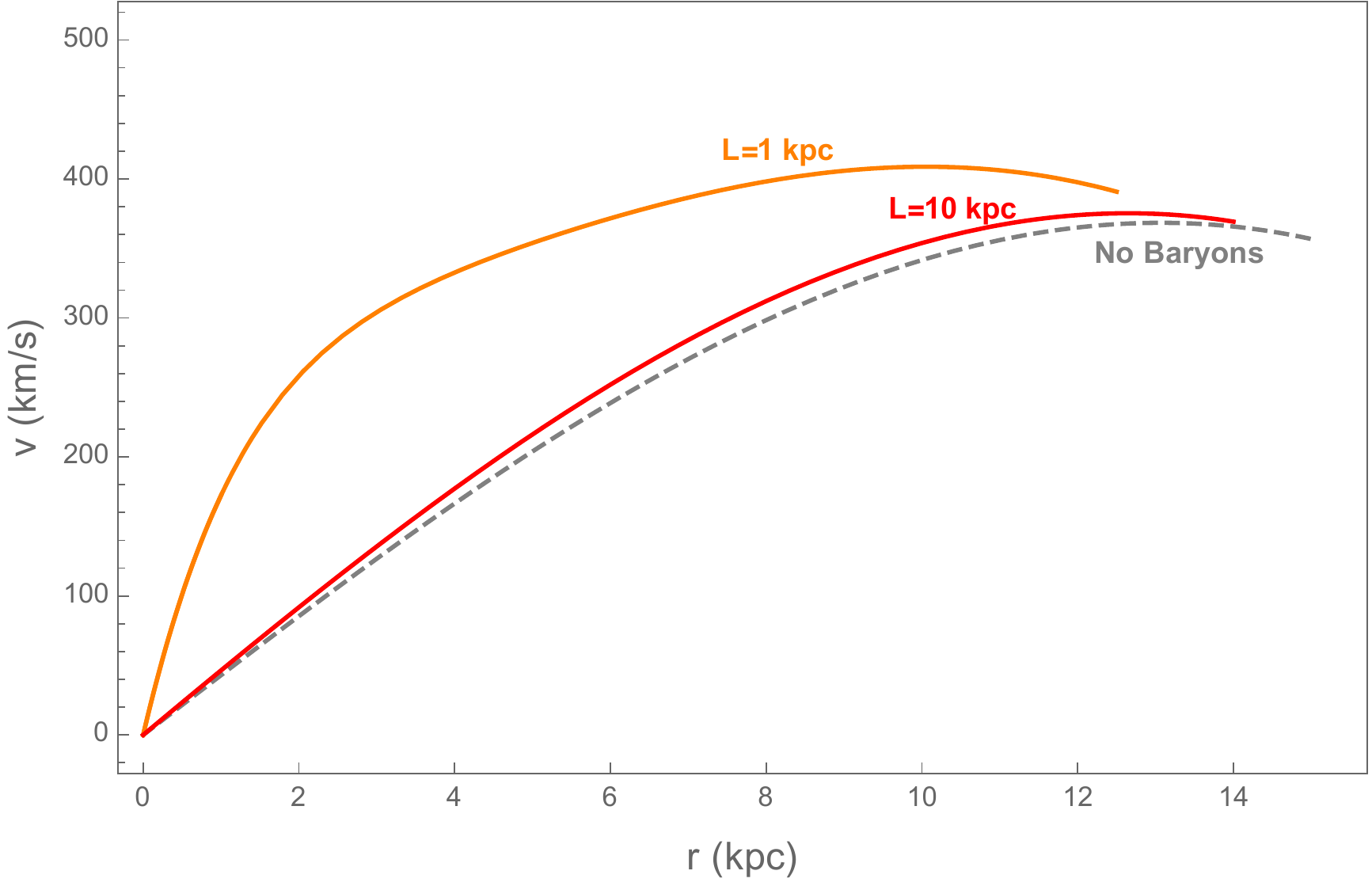} 	
  \caption{{\small{Rotation curves for configurations without baryons (grey dashed line), and baryon components with~$L=1$ (orange) and~$10$~kpc (red).}}}
\label{rotn curve}
\end{figure}

\begin{figure}
\centering
	\includegraphics[width=8cm]{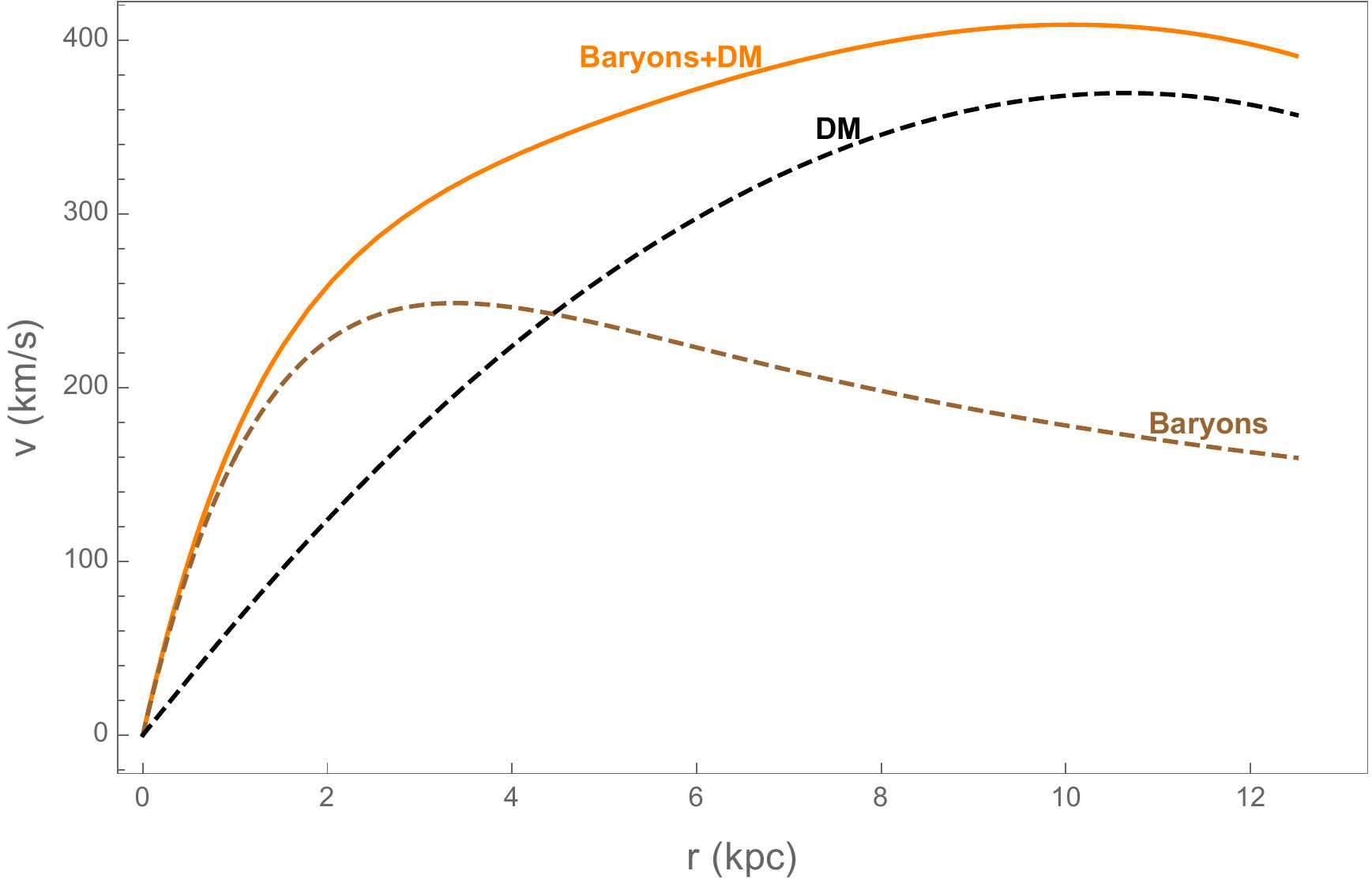}
		\includegraphics[width=8cm]{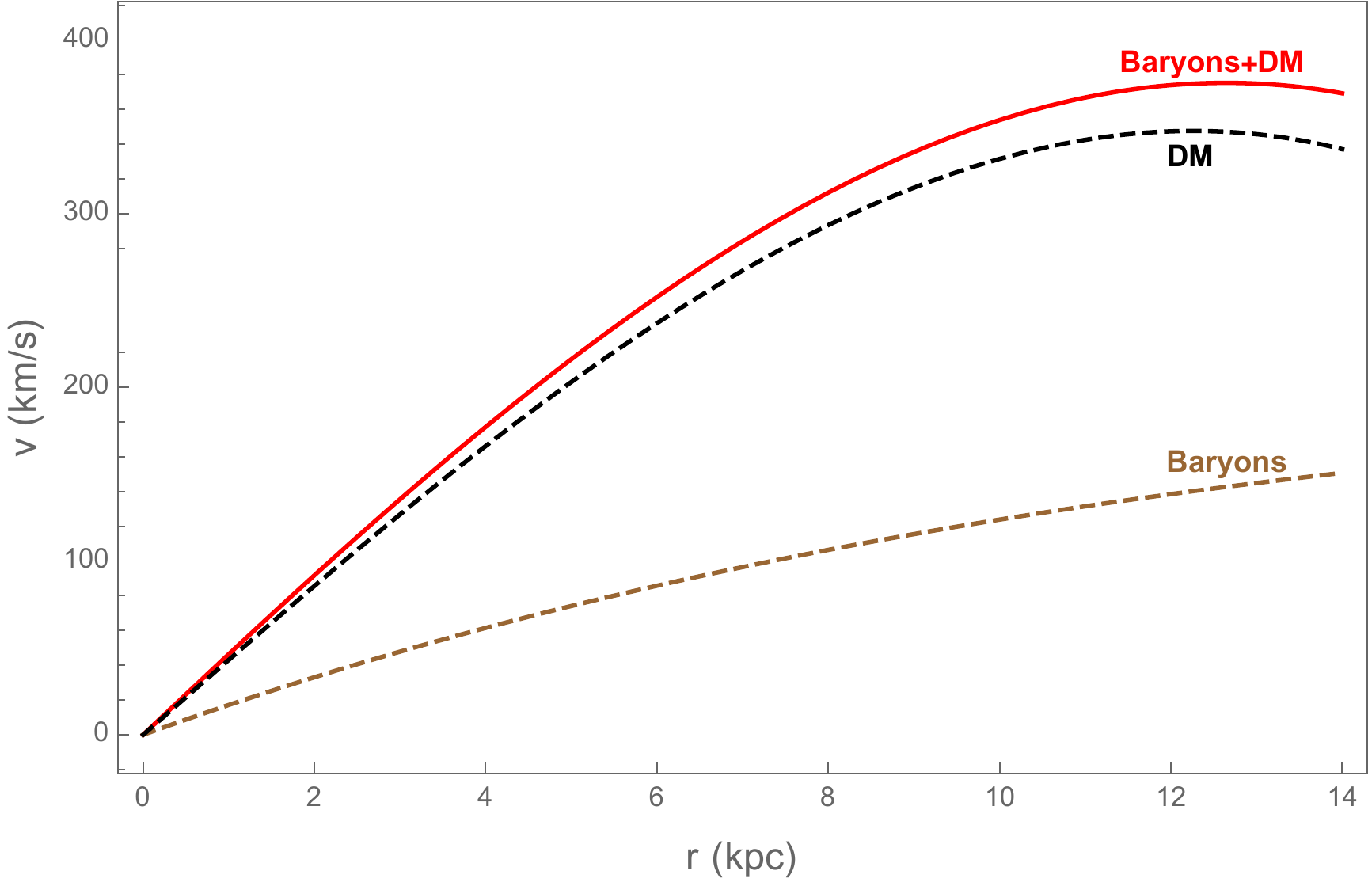}
  \caption{{\small{Individual contributions to the rotation curves of Fig.~\ref{rotn curve} for~$L=1$~kpc (Left) and~$L=10$~kpc (Right).}}} 
  \label{rotn curve contns}
\end{figure}

Figure~\ref{DM density prof} shows the DM density profile, obtained by solving the equation of hydrostatic equilibrium and Poisson's equation, in the presence of the baryonic component.
The solutions cannot be trusted close to the edge of the soliton, where effects from the DM envelope can no longer be neglected. As we can see, the soliton gets smaller for cuspy baryonic distributions, and the central dark matter density increases accordingly. Figure~\ref{rotn curve} shows the rotation curves in each case. We see a sharp rise of the rotation curve for small radii in the cuspy case, while the quasi-homogeneous case tracks the unperturbed case. The rotation curve in the absence of baryons is obtained by rescaling the central DM density by a factor of~1.2 to compensate for the missing mass fraction. Figure~\ref{rotn curve contns} shows the individual DM and baryon contributions to rotation curves, for~$L=1$~kpc (Left Panel) and~$L=10$~kpc (Right Panel).  
 
These results confirm the expectation that for a cored baryonic distribution, such as for LSB galaxy IC2574, the DM soliton maintains its homogeneous core. Correspondingly,
the rotation curve rises gradually, precisely as indicated by observations. In contrast, in the presence of a cuspy baryonic distribution, the superfluid soliton is rearranged into a cuspier profile,
and the rotation curve is found to rise sharply. However, it needs to be emphasized that the sharp rise is driven mainly by the gravitational field of baryons as illustrated in Fig.~\ref{rotn curve contns}.

This way of accounting for the diversity of rotation curves is somewhat similar to how this phenomenon is explained in SIDM scenario~\cite{Kamada:2016euw}. The difference in our case is that we do not rely on the homogeneity of the thermalized core, but merely on the presence of the central soliton (which is significantly smaller than the former). In fact, we made the case for the entire halo to be in equilibrium due to Bose enhancement of the scattering rates. Moreover, the thermal region was argued to be prone to fragmentation and tidal disruption. 

\section{Three-body Interacting Superfluid}
\label{3body sec}

In this Section, we explore the extent to which our results thus far are characteristic to the idea of DM superfluidity more generally, or are special to the theory with quartic self-interactions. We are particularly interested in
superfluidity with 3-body interactions, as this is the relevant case for the original DM superfluidity scenario~\cite{Berezhiani:2015bqa}. One of the main purposes of this consideration is to establish whether in the case of the partially thermalized halo the enhanced Bullet Cluster bound constrains the size of the superfluid soliton as tightly as for the quartic self-interaction. A key aspect of the quartic theory, as emphasized throughout this paper, is that the soliton size~$\ell$ is determined solely by the theory parameters~$m$ and~$\lambda$, and is therefore independent of the total mass.

The model we consider in this Section is one for the complex scalar field
\begin{equation}
\mathcal{L}_{\rm 3-body} = -|\partial\Phi|^2-m^2|\Phi|^2-\frac{g_3}{3}|\Phi|^6 \,.
\label{eq:3bodylagr}
\end{equation}
This choice is simply a matter of convenience, and the statements made in the non-relativistic limit are applicable to the real scalar field as well, as we have already pointed out in Sec.~\ref{superfluid_intro}. Similar to the quartic case, the scalar field under the consideration exhibits superfluidity upon condensation, albeit with a different equation of state; see, {\it e.g.},~\cite{Berezhiani:2015bqa}. 

\subsection{Scattering rate}

To analyze the thermalization rate in the virializing halo, we need to compute the rate of scattering of any given particle off the rest of the halo.  Although the latter is in the form of a non-relativistic gas, it is nevertheless convenient to imagine it as a condensate. Such a treatment is applicable when the scattering velocity exceeds the condensate sound speed~\cite{Berezhiani:2020umi}. In this case, the condensate description replaces the 3-body encounter by the process of phonon radiation. 

A particle, belonging to the same species as the condensate particles, propagates within it only as a phonon. As such, the rate of interactions we are interested in corresponds to the rate of hard phonons, with momentum~$k\gg mc_s$, radiating phonons
\begin{flalign}
&\pi\rightarrow  \pi+\pi\,; \nonumber\\
&\pi\rightarrow \pi+\pi+\pi\,;\nonumber\\
&\ldots 
\end{flalign}
Moreover, since the condensate description of the scattering target is merely a trick, one would have to include the Bose-enhancement factor by hand due to the fact that phonons are produced in already occupied states.

Another simplifying assumption we can make is to consider the incoming particle to be of a different species. Denoting this particle by~$\chi$, we can deform the theory~\eqref{eq:3bodylagr}, by retaining the only relevant interaction responsible for the supersonic scattering of interest:
\begin{equation}
\mathcal{L} = -|\partial \Phi|^2-m^2|\Phi|^2-3g_3|\Phi|^4|\chi|^2-|\partial \chi|^2-m^2|\chi|^2 \,.
\label{eq:3bodychi}
\end{equation}
It goes without saying that we will have to adjust the Bose-enhancement and symmetry factors appropriately by hand.

In this language, the computation of the thermalization rate via 3-body scatterings boils down to 
\begin{flalign}
&\chi\rightarrow  \chi+\pi\,; \nonumber\\
&\chi\rightarrow \chi+\pi+\pi\,;\nonumber\\
&\ldots
\end{flalign}
Here~$\pi$ is the high-momentum phonon, with the following Lagrangian
\begin{eqnarray}
\nonumber
\mathcal{L}_\pi & = & \frac{1}{2} \dot{\pi}^2-\frac{1}{2}\left(\frac{\vec{\nabla}^2}{2m} \pi\right)^2 -6\sqrt{2}g_3 V^3|\chi|^2\frac{\dot{\pi}}{\sqrt{-\vec{\nabla}^2}}-3{g_3 V^2}|\chi|^2\left[3\left(\frac{\dot{\pi}}{\sqrt{-\vec{\nabla}^2}}\right)^2+\left(\frac{\sqrt{-\vec{\nabla}^2}}{2m}\pi\right)^2\right]  \\
& & + \ldots
\end{eqnarray}
where~$V=\sqrt{\rho/2m^2}$ denotes the order parameter of the condensate, and ellipses stand for higher-order vertices for the phonon field. Notice that within~\eqref{eq:3bodychi} the process of radiating two phonons from a single vertex is physical due to the presence of 3-body interactions, which would not be the case if we had~$|\Phi|^2|\chi|^2$ instead. On the other hand, the rate of radiating more than two phonons vanishes to leading order in~$k\gg mc_s$. The rates of radiating a single and double phonon are given respectively as
\beq
\Gamma_{\chi\rightarrow\chi+\pi}=\frac{9g_3^2\rho^3}{2^{6}m^7 \pi} v\,; \qquad  \Gamma_{\chi\rightarrow\chi+\pi+\pi}=\frac{9g_3^2\rho^2}{2^{8}m^3 \pi^4} v^4\,,
\eeq
where we have evaluated the phase space integral of~$\Gamma_{\chi\rightarrow\chi+\pi+\pi}$ following the analysis of~\cite{Acanfora:2019con,Caputo:2019cyg}. 

Putting all the ingredients together, the scattering rate of a given particle off the rest of the halo can be computed as
\begin{equation}
\Gamma=\frac{1}{3!}\Big(\left(1+\mathcal{N}\right)\Gamma_{\chi\rightarrow\chi+ \pi}+\left(1+\mathcal{N}\right)^2\Gamma_{\chi\rightarrow\chi+\pi+\pi}\Big)= \mathcal{N}^2\frac{9g_3^2\rho^2}{2^{10}m^3 \pi^4} v^4\,.
\label{eq:rate}
\end{equation}
The Bose-enhancement factor~$\mathcal{N}$ has been inserted since, before thermalization, the DM phase space is one for the virialized gaseous medium. The~$1/3!$ factor has been introduced to take into account that final particles are three identical bosons in reality. It is worthwhile to point out that both channels give parametrically identical contributions due to the bosonic enhancement factor~$\mathcal{N}\gg 1$, where we have assumed the velocity dispersion to be of order~$v$ as before. In the absence of the degeneracy~$\mathcal{N}\ll 1$, the two-phonon channel would be the dominant one; in other words, it would have been more probable for the incoming particle~$\chi$ to transfer its momentum to two constituents of the condensate at a time.

\subsection{Thermal radius}

The condition delineating the region within which degenerate DM is in thermal equilibrium is obtained by combining~\eqref{eq:conditionRT} and~\eqref{eq:rate}:
\begin{equation}
\Gamma t_g=\mathcal{N}^2 \frac{9g_3^2}{4(4\pi)^4 m^3} \rho^2(R_T) v^4(R_T) t_g=1 .
\label{eq:ratetg}
\end{equation}
Here,~$v=\sqrt{\frac{GM(r)}{r}}$ is the orbital velocity of DM particles, while~$M(r)$ is the enclosed mass within radius~$r$. The density profile that enters~\eqref{eq:ratetg} is the NFW profile~\eqref{eq:NFW}.

As we can see from~\eqref{eq:ratetg}, the thermal radius~$R_T$ is a non-trivial function of the self-coupling~$g_3$ and the particle mass~$m$. However, the behavior simplifies both within the core of the halo ($R_T\ll r_s$) as well as in the outskirts ($R_T\gg r_s$):
\begin{equation}
R_T\simeq r_s\left(\frac{9\rho_{\rm NFW}^3}{32 Gr_s^2}\frac{g_3^2}{m^{11}} t_g\right)^\gamma\,,  \qquad \text{with}~~~~\gamma = \begin{cases} \frac{1}{5} & \mbox{for } R_T\ll r_s\,;  \\ \frac{1}{11} & \mbox{for } R_T\gg r_s \,. \end{cases}
\end{equation}
Because the density profile scales as~$r^{-3}$ in the outskirts, stronger self-interactions (or larger Bose enhancement factor) are needed to compensate for the drop in density.  Moreover, the case $R_T\gg r_s$ shows that thermalization in the outer regions depends strongly only on the mass of the DM candidate, and is quite insensitive to the strength of the interactions, due to the squared dependence of the interaction rate on the Bose enhancement factor.

Finally, it is possible to derive the Bullet Cluster bound in the case of three-body interacting DM. Using~\eqref{eq:rate} in~\eqref{eq:RateBull}, we find that the following parameter strip is ruled out:
\begin{equation}
4\left(\frac{m}{ \text{eV}}\right)^{11/2} \text{keV}^{-2}\lesssim g_3\lesssim6\times10^3\left(\frac{m}{ \text{eV}}\right)^{11/2} \text{keV}^{-2}   , \qquad\mbox{for}~~~~ m\ll \text{eV}.
\label{eq:bullet}
\end{equation}
As before, the left inequality comes from the Bullet Cluster bound for the case of significant fraction of the DM being in the gaseous form with degenerate phase space, while the right inequality is saturated when clusters reach complete thermal equilibrium.

Without Bose enhancement, on the other hand, we find the excluded region:
\begin{equation}
g_3\gsim 200 \left(\frac{m}{\text{eV}}\right)^{3/2}\text{keV}^{-2}\,,\quad \mbox{for}~~m\gg \text{eV} ~~\text{or clusters in thermal equilibrium\,.}
\end{equation}
Here, the assumptions made about the DM properties within clusters are identical to the ones of Sec.~\ref{bullet}, where we dealt with 2-body interactions.

\subsection{Coherence length}

We now have at our disposal all the ingredients to study the consequences for the Jeans scale~$\ell$. The dispersion relation is given by~\eqref{eq:dispersion}, as before, but with a modified sound speed:
\beq
c_s^2=\frac{g_3 \rho^2}{4m^6}\,.
\eeq
It follows from the dispersion relation that the Jeans length is given by
\begin{equation}
\ell=\left(\frac{\pi g_3\rho}{4 G m^6}\right)^{1/2}.
\label{eq:3bJS}
\end{equation}

If we apply the Bullet Cluster bound, assuming partial thermalization or significant left-over degenerate gaseous DM component in clusters, the largest possible Jeans length follows from the saturation of the left-hand inequality of~\eqref{eq:bullet}, which leads to
\beq
\ell\lesssim 10^{-4}\left( \frac{m}{\rm eV}\right)^{-1/4}{\rm kpc}\,.
\eeq
In other words, one would need to have~$m\lesssim 10^{-16}\,{\rm eV}$ in order to accommodate~$\sim {\rm kpc}$ coherence length. The situation is somewhat similar to the case of 2-body interactions, which means that the modification of the type of the self-interaction does not help. Even in the case at hand, one needs to rely on the complete thermalization of cluster halos.

The numerical analysis of the dependence of $R_T$ and $\ell$ on parameters of the theory demonstrates the same hierarchy between these length scales as in the case of quartic interactions. This similarity for 2 and 3-body interacting superfluids implies that the thermal core must be prone to fragmentation in the second case, as it is in the first one. 

This might be connected to the following instructive ``dictionary'' between the two theories:
\begin{equation}
\frac{P_2}{\sqrt{\Gamma_2}}\simeq  \frac{P_3}{\sqrt{\Gamma_3}} \qquad \text{if}\qquad m\ll\text{eV}. 
\label{PvsTh}
\end{equation}
Here~$\Gamma_2$ and~$\Gamma_3$ are given by~\eqref{eq:csRate} and~\eqref{eq:rate}, respectively, while
\begin{equation}
P_2=\frac{g_2 \rho^2}{8m^4} \qquad {\rm and} \qquad P_3=\frac{g_3\rho^3}{12m^6}
\end{equation}
are the equation of state for 2- and 3-body interacting theories. In the above,~$g_2$ is the coupling constant of the quartic version of~\eqref{eq:3bodychi}, {\it i.e.}, the latter is compared to the complex scalar field theory
\beq
\mathcal{L}_{\rm 2-body}=-|\partial \Phi|^2-m^2|\Phi|^2-\frac{g_2}{2}|\Phi|^4\,.
\eeq

The density profile of the self-gravitating superfluid soliton is obtained as before by solving the equation for hydrostatic equilibrium and Poisson's equation. In this case, the resulting equation is equivalent to the~$n=1/2$ Lane-Emden equation. Even though the analytic solution is unknown for this system, the function
\begin{equation}
\rho(r)\simeq \rho_0\cos^{1/2}\left(\frac{\pi}{2} \frac{r}{R}\right)\,, \qquad \text{with}\quad R=R_0\sqrt{\frac{g_3 \rho_0}{32\pi G m^6}} \,,
\label{eq:profile}
\end{equation}
provides a good fit to the profile~\cite{Berezhiani:2015bqa}. Here~$R_0\simeq 2.75$ is the first zero of the exact solution to the Lane-Emden equation. Furthermore, the size of the soliton is related to the Jeans scale as
\begin{equation}
R\simeq 0.3\, \ell\,.
\end{equation}
Thus, in the case of 3-body self-interactions, the diameter of a stable soliton is roughly 60\% of the Jeans length, unlike the quartic theory~\eqref{soliton_profile} where the diameter was equal to the latter. 
It is important to note that the tidal disruption of non-central solitons is expected to proceed even more efficiently than for the quartic case, due to the fact that~$\ell \propto \sqrt{\rho}$ for 3-body interactions.

One might wonder if the situation could be remedied by decoupling the dynamics of the superfluid phase from the process of thermalization. In other words, one could think of introducing a potential:
\begin{equation}
\mathcal{L}\supset -\frac{g_2}{2}|\Phi|^4-\frac{g_3}{3}|\Phi|^6 \,,
\end{equation} 
and have a system where 3-body processes drive thermalization, while 2-body interactions regulate the dynamics of the superfluid phase. In such a system, the relaxation rate and total pressure would read:
\begin{equation}
\Gamma_\text{rel}=\Gamma_2+\Gamma_3\,; \qquad  P=P_2+P_3 \,.
\end{equation} 
As long as DM is degenerate, we may recast~\eqref{PvsTh} as
\begin{equation}
\sqrt{\frac{\Gamma_3}{\Gamma_2}}\simeq\left(\frac{g_3\rho}{g_2m^2}\right)\simeq\frac{P_3}{P_2}\,.
\end{equation} 
We can see from this relation that if one type of interaction sustains the soliton, it must drive thermalization as well. Therefore, this ``hybrid'' solution inevitably fails.\footnote{Notice that this would not have necessarily been the case in the absence of Bose enhancement factors in interaction rates. In that case,~$\frac{P_3}{P_2}\sim\left(\frac{\text{ eV}}{m}\right)^2\sqrt{\frac{\Gamma_3}{\Gamma_2}}$, and for~$m\gg \text{eV}$ it is possible to make 3-body interactions driving thermalization, while the pressure mainly being sourced by 2-body interactions. However, the mass range that would facilitate the removal of bosonic enhancement for scattering is incapable of producing large superfluid coherence lengths, which is of our interest.}

As  demonstrated in~\cite{Berezhiani:2021rjs}, and depicted in Fig.~\ref{fig:Fig1}, the parameter space of the quartic theory required for the formation of macroscopic superfluid solitons does not entail a significant pressure at matter-radiation equality, to cause a tension with cosmological observations on large scales. The situation changes in the case of 3-body interactions, as the superfluid equation of state tends to become relativistic by matter-radiation equality for such theories~\cite{Berezhiani:2015bqa}, unless an additional mechanism is invoked for softening the pressure at high densities. In the Appendix, we provide a method for achieving this.

\section{Summary}

In this work we have provided a refined analysis of DM halo substructure in the context of the sub-eV, self-interacting DM scenario capable of forming a superfluid upon Bose-Einstein condensation. 

In previous work~\cite{Berezhiani:2021rjs}, it had been argued that the formation of a superfluid phase in the inner regions of galaxies would leave a significant un-condensed, degenerate fraction of DM in the outskirts of halos. Because self-interactions would not be efficient enough to rearrange the disordered phase space, DM self-interactions in those regions were claimed to be Bose-enhanced, which translated into an enhancement of the DM interaction rate.  This led to a revised bound on the cross section inferred from the Bullet Cluster measurements, which would in turn constrain the parameter space in such a way as to impede the superfluid formation with a coherence length greater than~$\sim {\rm kpc}$. As a result, even if a significant fraction of the halo reached nearly thermal equilibrium, this thermalized region would be prone to fragmentation into solitonic superfluid islands of the corresponding size, due to the Jeans instability.

In this work, we have exploited a possible loophole for lifting the tightened Bullet Cluster constraint by means of fully thermalizing the halo. In this case, the entire halo would undergo the superfluid phase transition and fragmentation into solitonic patches. If, in the process, there was not a significant DM fraction left over in the gaseous form, then the cross-section constraint from cluster mergers would revert to the conventional one, and allow for the formation of superfluid islands of tens of kpc size. 

We have further demonstrated that all of these solitons would undergo tidal disruption, with the exception of the central soliton. The tidally disrupted superfluid islands are expected to behave as virialized superfluid streams. In other words, in the most interesting region of parameter space, DM halos are entirely in the fluid form, with a coarse-grained density distribution similar to NFW profile. Furthermore, the density profile of the central superfluid soliton is influenced by the distribution of baryons in the core, in particular whether the latter is cored or cuspy, with interesting ramifications for the diversity of rotation curves of dwarf galaxies.

In the last part, we have studied the extent to which the story changes with more general self-interaction potential. In particular, we have compared the results of 3-body interactions with the original quartic case. We found that this does not improve the picture. In fact it brings up an additional complication, namely that non-renormalizable interactions for sub-eV scalar DM tends to lead to a relativistic equation of state at matter-radiation equality. In the Appendix, we demonstrate how to ameliorate this issue by integrating in a heavy particle that recasts the theory into a renormalizable one.
 
Going back to the simplest theory with quartic interactions, the upshot of our analysis is that the end state of superfluid DM in galactic halos is clearly quite complex. The interplay of thermalization, fragmentation and tidal disruption leaves the superfluid DM in a turbulent fluid state, featuring streams of superfluid debris. It will be fascinating to see whether these expectations are borne out by numerical simulations, and whether the latter can unveil further, unexpected complex phenomena of the superfluid scenario.     

\section*{Acknowledgements}

The work of J.K. is supported in part by the DOE (HEP) Award DE-SC0013528.

\appendix 
\section{Softening Pressure at Matter-Radiation Equality}
\renewcommand{\theequation}{A-\Roman{equation}}
\setcounter{equation}{0}

In this Appendix we discuss a possible mechanism to alleviate the issue of relativistic equation of state at matter-radiation equality, for the superfluid scenario with 3-body interactions discussed in Sec.~\ref{3body sec}.

The requirement for the interaction pressure due to 3-body interactions being sufficiently small at matter-radiation equality entails:
\begin{equation}
\left.\frac{P}{\rho}\right\vert_{\text{equality}}=\frac{g_3\rho^2_{\text{eq}}}{12m^6}\ll1 \,.
\label{eq:pressure3b}
\end{equation}
Substituting~$\rho_\text{equality}\simeq 0.4\text{ eV}^4$ and using~\eqref{eq:profile}, we can convert the bound on the pressure to a restrictive bound on the size of a superfluid soliton:
\beq
R\ll 0.1 \text{ kpc}\,.
\label{Rbound}
\eeq
In this work, we have not encountered any advantage of generalizing the DM self-interaction potential beyond the quartic one. However, due to the relevance of 3-body interactions for certain extensions of DM superfluidity~\cite{Berezhiani:2015bqa}, it is instructive to show how  the aforementioned problem can be tackled by embedding the sextic model~\eqref{eq:3bodylagr} into a UV-complete model which will behave as the quartic theory at matter-radiation equality.

For this purpose, consider the following two-field theory
\begin{equation}
\mathcal{L}=-\frac{1}{2}(\partial \chi)^2-\frac{1}{2}M^2\chi^2-|\partial \phi|^2-m^2|\phi|^2-\frac{g}{2}\chi^2|\phi|^2-\mu \chi |\phi|^2-\frac{\delta}{2}|\phi|^4\,.
\label{eq:2fieldLagr}
\end{equation}
This is an extension of the simple quartic model by introducing a coupling between the DM field~$\phi$ and an additional real boson~$\chi$. 
Assuming that~$\chi$ is heavy enough to be integrated out in the regime of interest, we can do so and truncate the resulting effective field theory of~$\phi$ at leading-order in derivatives
\begin{equation}
\mathcal{L}=-|\partial\phi|^2-m^2|\phi|^2+\frac{1}{2}\frac{\mu^2|\phi|^4}{M^2+g|\phi|^2}-\frac{\delta}{2}|\phi|^4\,.
\label{eq:1field}
\end{equation}
If one is going to expand the Lagrangian~\eqref{eq:1field} in powers of~$\phi$,  the effective theory thus obtained would contain terms that mediate 2-body as well as higher-order interactions, with the latter being suppressed by powers of~$M^{-2}$. However, if~$\phi$ acquires a vacuum expectation value~$V$, the mass of~$\chi$ will get shifted as:
\begin{equation}
m_\chi^2=M^2+gV^2\,.
\end{equation} 
Expanding the Lagrangian~\eqref{eq:1field} around~$M$ in this case is legitimate only if~$M^2\gtrsim gV^2~$.

Therefore, to understand the complete picture, we have to study the fully re-summed Lagrangian~\eqref{eq:1field}. In particular, if~$\phi$ particles are in a condensed phase, we may introduce the following classical solution to the equation of motion of~$\phi$:
\begin{equation}
\phi_0=V{\rm e}^{{\rm i}\omega t}\,;  \qquad \omega^2=m^2+\left(\delta-\frac{\lambda}{1+\frac{V^2}{\Lambda^2}}\right) V^2+\frac{\lambda V^4}{2\Lambda^2\left(1+\frac{V^2}{\Lambda^2}\right)^2}\,,
\label{eq:class}
\end{equation}
 where~$\lambda\equiv\frac{\mu^2}{M^2}$ and~$\Lambda^{-2}\equiv\frac{g}{M^2}$.
The pressure generated by this ground state solution reads
\begin{equation}
P=\left(\delta -\frac{\lambda}{1+\frac{V^2}{\Lambda^2}}\right)\frac{V^4}{2}+\frac{\lambda}{2\Lambda^2}\frac{V^6}{(1+\frac{V^2}{\Lambda^2})^2} \,.
\end{equation}

From this, we note that two limiting behaviors of the pressure may be found, depending on whether or not the mass of~$\chi$ is dominated by the shift~$gV^2$:
\begin{equation}
P  \simeq  \begin{cases} \frac{\delta V^4-\lambda \Lambda^4}{2} &\mbox{for }  \Lambda^2\ll V^2\,;  \\ \frac{(\delta-\lambda) \rho^2}{8 m^4}+\frac{\lambda  \rho^3}{8\Lambda ^2m^6} &\mbox{for } \Lambda^2\gg V^2\,.\end{cases}
\label{eq:pressure2}
\end{equation}
This relation shows there are three different scenarios, depending on the relative magnitude of~$\delta$ and~$\lambda$:
\begin{itemize}
\item If~$\delta \gg \lambda$, 2-body interactions stabilize solitons, and the model reduces to the original quartic model. In this case, the pressure that sustains kpc-size solitons does not spoil the formation and evolution of cosmological structures.
\item If~$\delta \ll \lambda$ and~$\Lambda^2\gg V^2$, long-wave perturbations of the condensate are always tachyonic. On the other hand, if~$\Lambda$ is sub-leading, the interaction pressure is positive for~$\frac{V}{\Lambda}>\frac{\lambda}{\delta}$. As we may see, had we started with a field~$\phi$ without contact self-interactions, the pressure generated by the condensate would have been negative, independently of the relative magnitude of~$V$ and~$\Lambda$, similarly to what takes place for an axion condensate~\cite{Guth:2014hsa}. 
\item If~$\delta\simeq \lambda$, the two-body contribution to the pressure vanishes in the limit~$\Lambda^2\gg V^2$, and the model reduces to the model~\eqref{eq:3bodylagr}. On the other hand, for~$\Lambda^2\ll V^2$, the pressure generated by the condensate is the same one we would have had for a~$\lambda |\Phi|^4$ condensate.
\end{itemize}
 Let us focus on the last case and consider the following condition on the parameters of the theory:
\begin{equation}
\rho_{\text{galaxy}}\ll\frac{m^2 M^2}{g}\ll\rho_{\text{eq}}.
\label{eq:window}
\end{equation}
Thanks to the left inequality, the theory reduces to~\eqref{eq:3bodylagr} for typical galactic densities, while the right inequality leads to condensates which are sustained by 2-body interactions at matter-radiation equality. In particular, we will exploit this last property to alleviate the excess of interaction pressure at the time of recombination. 

To connect the physics of the galactic condensate to its early universe counterpart, we note that the effective coupling~$g_3$ appearing in~\eqref{eq:3bodylagr} is connected to the parameters of the UV-complete theory by:
\begin{equation}
 g_3=\frac{3\lambda}{2\Lambda^2}=g\frac{\mu^2}{M^4}\,.
\end{equation}
Combining this relation with~\eqref{eq:pressure2} and~\eqref{eq:3bJS} leads to the following revision of~\eqref{Rbound} for the bound on the size of solitons: 
\begin{equation}
\frac{R}{\text{kpc}}\ll 0.6\sqrt{g \frac{\text{ eV}^4}{m^2M^2}}\,,
\label{eq:pressure}
\end{equation}
Let us stress that the pressure we considered here is given by the first line of~\eqref{eq:pressure2} since we are assuming to be in the regime~\eqref{eq:window}. Moreover, notice that this bound is independent of~$\mu$, which only enters in the Bullet Cluster bound. Here, we have used~$\rho_\text{galaxy}\simeq 10^{-25}\,\text{g}/\text{cm}^{-3}$ as the mean density of a soliton to rewrite the pressure in terms of~$R$.  Choosing~$R\simeq 10$ kpc, then~\eqref{eq:pressure} implies~$\frac{m^2M^2}{g}\ll 10^{-3}\,{\rm eV}^4$, which lies within the window~\eqref{eq:window}.

\end{document}